\documentstyle[seceq,preprint]{jpsj}
\def\runtitle{Magneto-Coulomb Oscillation in Ferromagnetic Single Electron Transistors}
\def\runauthor{H.{\sc Shimada}, K. {\sc Ono} and Y. {\sc Ootuka}}
\title
{
Magneto-Coulomb Oscillation \\ in Ferromagnetic Single Electron Transistors
}

\author
{
Hiroshi {\sc Shimada}\footnote{E-mail: crcsim@crc.u-tokyo.ac.jp}, Keiji {\sc Ono} and Youiti {\sc Ootuka}, 
}
\inst
{
Cryogenic Center, The University of Tokyo\\
2-11-16 Yayoi, Bunkyo-ku, Tokyo 113\\
}

\recdate
{
}

\abst{
The mechanism of the magneto-Coulomb oscillation in ferromagnetic single electron transistors (SET's) is theoretically considered. Variations in the chemical potentials of the conduction electrons in the ferromagnetic island electrode and the ferromagnetic lead electrodes in magnetic fields cause changes in the free energy of the island electrode of the SET. This is a plausible origin of the conductance oscillation of the SET in sweeping an applied magnetic field.
}

\kword
{
single electron charging effect, Coulomb oscillation, ferromagnet, SET
}

\begin{document}
\sloppy
\maketitle

\section{Introduction}
There have been intensive studies on the single electron charging effects \cite{AvrLik,SCT} last several years. In experimental works on metallic devices, however, the materials investigated have been restricted to a few kinds of metals such as Al, for example. Combination of materials of different characteristic properties, such as normal-conducting metals, superconductors, ferromagnetic metals, could add more fertile aspects to the physics of the single electron devices.

 Recently the present authors have found several novel phenomena in ferromagnetic single electron devices. \cite{Ono1,Ono2} One of them is the magneto-Coulomb oscillation in a ferromagnetic single electron transistor (SET). \cite{Ono2} The ferromagnetic SET in ref. 4 is composed of a Co island electrode and Ni lead electrodes. A conductance oscillation similar to the Coulomb oscillation \cite{SCT} was observed in sweeping an applied magnetic field. In this paper we discuss the mechanism of this phenomenon with supplementing experimental results on a Ni/Co/Ni ferromagnetic SET.

The essential origin of the magneto-Coulomb oscillation in the ferromagnetic SET is thought of to be noticeable change in the chemical potential of an isolated ferromagnet in magnetic fields. This causes different influences on the operation of the SET depending on whether its island electrode is composed of a ferromagnetic metal or the lead electrode is ferromagnetic. We describe the magneto-Coulomb oscillation based on the linear response theory of the Coulomb oscillation \cite{Beenakker,KulShek} taking into consideration the behavior of the ferromagnetic component of the SET in magnetic fields. Bearing the experimental situation in ref. 4 in mind, we consider the case in the classical regime where average spacing of the one-electron energy levels (${\it \Delta} \epsilon$) in the island electrode is much smaller than the thermal energy of the electrons in the system (${\it \Delta}\epsilon \ll k_{\rm B}T$) and the energy distribution of the electrons in the island electrode is well described by the Fermi-Dirac distribution. \cite{Kubo} We also assume that the tunnel resistance of the tunnel junctions are sufficiently larger than the quantum resistance, $R_{\rm Q}(=h/e^{2})$, and the orthodox theory of the single-electron tunneling \cite{AvrLik} is applied to the case without considering the quantum charge fluctuations or the renormalization of the charging energy.

The paper is organized as follows: 
In $\S$2 magnetic-field-induced variations in the chemical potential and in the work function of the ferromagnetic island electrode are examined. Here we regard the island electrode as an isolated small grain of a ferromagnetic metal. 
In $\S$3, effect of the magnetic field for the ferromagnetic lead electrode and its influence on the operation of the SET are considered. 
Based on the consideration in these sections, the thermodynamic potential of the ferromagnetic island electrode is constructed in $\S$4. This quantity dominates the character of the transport through the SET in the linear response regime. \cite{Beenakker,KulShek} 
In $\S$5, we examine the magneto-Coulomb oscillation in the ferromagnetic SET in some detail adopting a simple model for the ferromagnet.
More realistic consideration on the ferromagnet and corrections to the results in $\S$2-5 are presented in $\S$6. 
Experimental results on the magneto-Coulomb oscillation in the Ni/Co/Ni SET which supplement the result in ref. 4 are presented in $\S$7 and are discussed in $\S$8. Possible applications of this phenomenon is discussed in $\S$9, and the conclusion is described in $\S$10.

\section{Change in the Work Function of the Ferromagnetic Island Electrode}
The behavior of the ferromagnetic metal in a magnetic field has many aspects, and it has not been sufficiently understood until now. Therefore, in order to describe the essence of the mechanism of the magneto-Coulomb oscillation in the ferromagnetic SET, we adopt for a moment the simplest model for a ferromagnetic metal, that is, a degenerate free electron system that has spin-dependent density of states, and ignore the exchange-correlation effect for excitations. We also ignore effect of the orbital angular moments of the electrons. These contributions will be discussed in $\S$ 6. Moreover, for the sake of simplicity, we assume, in the following discussions, that the ferromagnet under consideration has a single domain. The effect of the multi-domain structure of the actual ferromagnet will be discussed later.

Suppose a ferromagnet in an applied magnetic field, $H$, with the direction of its magnetization parallel to $H$. Because of the Zeeman effect, the energy band of the majority-spin electrons shifts to the lower energy, in comparison with the case in zero magnetic field, by $\delta(H)=g\mu_{\rm B}H/2$, where $g$ is the electronic g-factor and $\mu_{\rm B}$ is the Bohr magneton. On the other hand, the energy band of the minority-spin electrons shifts to the higher energy by $\delta(H)$. In equilibrium, re-population between the majority-spin electrons and the minority-spin electrons occurs through spin-flip scatterings (Fig. 1). 
\begin{figure}
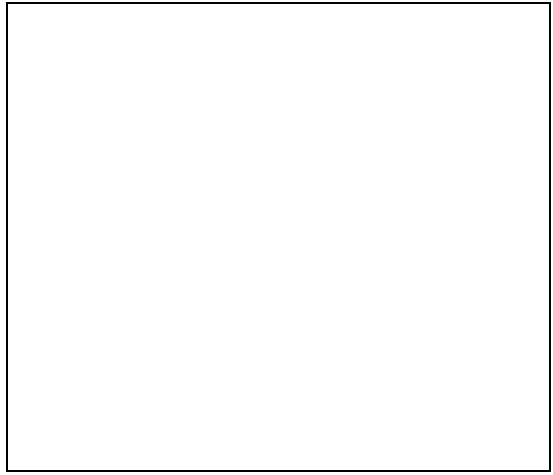

\figureheight{6cm}
\caption{Schematic energy band diagram of the majority($+$)- and the minority($-$)-spin conduction electrons in a ferromagnetic metal in zero magnetic field (broken line) and in a finite magnetic field $H$ (solid line). 2${\it \Delta}$ denotes the exchange splitting of the two subbands in zero magnetic field. $\delta(H)$ indicates the magnitude of the shift of the bands in $H$. $\zeta(0)$ and $\zeta(H)$ respectively indicate the chemical potentials of the conduction electrons in zero magnetic field and in $H$. $W(0)$ and $W(H)$ means the work functions in zero magnetic field and in $H$, respectively. $-{\cal E}_{c}$ is the midpoint between the bottoms of the two conduction subbands measured from the vacuum level.}
\label{fig:1}
\end{figure}
Now we consider a ferromagnetic island electrode with electrons of a fixed number $N$. $N$ is a sum of the number of the majority-spin electrons, $N_{+}$, and that of the minority-spin electrons, $N_{-}$. The equivalence of the total number of electrons both in zero magnetic field and in magnetic field $H$ is expressed as follows: 
\begin{eqnarray}
\lefteqn{
\int_{-{\it \Delta}}^{\infty}\rho_{+}(\epsilon)f(\epsilon-\zeta(0)){\rm d}\epsilon
+\int_{+{\it \Delta}}^{\infty}\rho_{-}(\epsilon)f(\epsilon-\zeta(0)){\rm d}\epsilon
} \nonumber \\
& =\int_{-{\it \Delta}-\delta(H)}^{\infty}\rho_{+}(\epsilon+\delta(H))f(\epsilon-\zeta(H)){\rm d}\epsilon
+\int_{+{\it \Delta}+\delta(H)}^{\infty}\rho_{-}(\epsilon-\delta(H))f(\epsilon-\zeta(H)){\rm d}\epsilon.
\end{eqnarray}
Here we take, as the origin of the energy of an electron, the midpoint between the bottoms of the conduction subbands of the majority- and the minority-spin electrons, and name the difference of the energy between this point and the vacuum level of an electron ${\cal E}_{\rm c}$ (Fig. 1). Throughout this paper, the vacuum level is defined at infinity where magnetic field and the electrostatic potential is considered to be zero. \cite{WF} In the above expression $\rho_{\pm}(\epsilon)$ indicate the densities of states of the majority($+$)- and the minority($-$)-spin electrons in zero magnetic field, and $f(x)$ is the Fermi-Dirac distribution function:
\begin{equation}
f(x)=\frac{1}{1+{\rm e}^{x/k_{\rm B}T}}.
\end{equation}
The symbol ${\it \Delta}$ denotes a half of the magnitude of the exchange splitting at the bottoms of the conduction subbands and $\zeta(H)$ is the chemical potential of the conduction electrons in a magnetic field $H$. \cite{CP} The work function in a magnetic field $H$ is expressed as $W(H)={\cal E}_{\rm c}-\zeta(H)$.

Equation (2.1) is rewritten to a balance equation for the re-population between the majority- and the minority-spin electrons in magnetic fields:
\begin{eqnarray}
\lefteqn{
\int_{-{\it \Delta}}^{\infty}\rho_{+}(\epsilon)
[f(\epsilon-\zeta(0))-f(\epsilon-\delta(H)-\zeta(H))]{\rm d}\epsilon
} \nonumber \\
& +\int_{+{\it \Delta}}^{\infty}\rho_{-}(\epsilon)
[f(\epsilon-\zeta(0))-f(\epsilon+\delta(H)-\zeta(H))]{\rm d}\epsilon
=0.
\end{eqnarray}
In ferromagnets such as 3$d$ transition metals, a condition, $k_{\rm B}T\ll\zeta-{\it \Delta}$, is satisfied at temperatures below room temperature, and the lower limits of the above integrals can be replaced by 0. In not extremely strong magnetic fields, a condition, $\delta(H)/\zeta\ll 1$, is satisfied and we expand the Fermi-Dirac function as the power of $\delta(H)/\zeta(0)$ and ${\it \Delta}\zeta(H)/\zeta(0)$ (${\it \Delta}\zeta(H)=\zeta(H)-\zeta(0)$). Up to the first order in these quantities, eq. (2.3) becomes
\begin{equation}
\int_{0}^{\infty}\frac{\rho_{+}(\epsilon)\{\delta(H)+{\it \Delta}\zeta(H)\}
-\rho_{-}(\epsilon)\{\delta(H)-{\it \Delta}\zeta(H)\}}{1+{\rm cosh}[\,(\epsilon-\zeta(0))/k_{\rm B}T\,]}
{\rm d}\epsilon
=0.
\end{equation}
This equation determines the change in the chemical potential ${\it \Delta}\zeta(H)$, and, therefore, that in the work function of the ferromagnet in magnetic fields. If the densities of states of the conduction electrons over an energy region of the order of $k_{\rm B}T$ around $\zeta$ is constant, eq. (2.4) gives a simple relation:
\begin{equation}
(\rho_{+}-\rho_{-})\delta(H)+(\rho_{+}+\rho_{-}){\it \Delta}\zeta(H)=0,
\end{equation}
where $\rho_{\pm}=\rho_{\pm}(\zeta(0))$. This gives the shift of the chemical potential:
\begin{equation}
{\it \Delta}\zeta(H)=-{\cal P}g\mu_{\rm B}H/2,
\end{equation}
where
\begin{equation}
{\cal P}=\frac{\rho_{+}-\rho_{-}}{\rho_{+}+\rho_{-}}.
\end{equation}
This quantity ${\cal P}$ denotes the spin polarization of the electronic densities of states at the Fermi energy of the ferromagnetic metal under consideration. It should be noted that this quantity is not the same one as measured in the spin-polarized tunneling experiment. The latter is weighted by the tunneling probability of each conduction band at the Fermi energy. \cite{Stearns,Maekawa,MesTed}

\section{Effect of the Magnetic Field for the Ferromagnetic Leads}
The chemical potential of the ferromagnetic lead electrodes also changes in magnetic fields, which causes another effect on the operation of the SET. In real experimental situations, the ferromagnetic leads are connected to non-magnetic metallic leads, which are connected to electronics, and the whole system is finally anchored to the ground somewhere in the circuit. In order to make the discussion clear, we assume a system depicted in Fig. 2 in the following discussions.
\begin{figure}
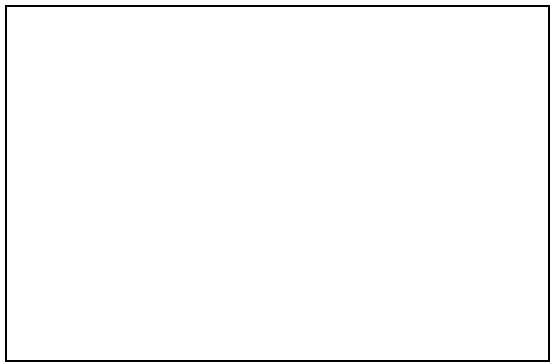

\figureheight{4.5cm}
\caption{The ferromagnetic SET. A and B denote a ferromagnetic island electrode and ferromagnetic lead electrodes, respectively. The ferromagnetic metals of A and B is not necessarily the same. C denotes a non-magnetic metallic lead. ${\cal C}$ and ${\cal C}_{\rm g}$ respectively indicate the capacitance of each tunnel junction and that of the gate capacitor. $V_{\rm g}$ is the electrostatic-potential-difference source for the gate and it also means the electrostatic-potential difference itself in the text.}
\label{fig:2}
\end{figure}

In general, there exists some difference in the electrostatic potential between two kinds of metals in contact, because the work functions of the two metals before contact are different in most of cases. Once they are in contact, electrons are transferred from the metal of the lower work function to the other, and an electric dipole layer is generated at their interface up to the depths around their screening lengths. The electric dipole layer makes their electrostatic potentials shift, and, therefore, changes the electrostatic-potential energies of electrons in the two metals so as to equate their {\it electrochemical} potentials (and the work functions) of both metals. Therefore {\it electrostatic} potentials of the ground, the non-magnetic leads and the ferromagnetic leads in Fig. 2 are considered to be different in general, although the {\it electrochemical} potential, $\omega$, and the work function, $W$, are kept constant over the combined system in equilibrium.

In magnetic fields, as describe in the previous section, the chemical potential of the ferromagnetic lead electrode changes by ${\it \Delta}\zeta(H)=-{\cal P}g\mu_{\rm B}H/2$. Assume ${\cal P}$ is negative, for example, in the ferromagnet under consideration. Then ${\it \Delta}\zeta(H)$ becomes positive. In this case electrons flow from the ferromagnetic leads to the non-magnetic leads through their interfaces, the charges of the electric dipole layers at the interfaces change, the electrostatic potential of the ferromagnetic leads is raised, and the electrostatic-potential energy of conduction electrons is lowered to compensate the change in its chemical potential. As a result, the contact potential difference, $V_{\rm c}$, changes by
\begin{eqnarray}
{\it \Delta} V_{\rm c}(H)&=&{\it \Delta}\zeta(H)/e \nonumber \\
&=&-{\cal P}g\mu_{\rm B}H/2e,
\end{eqnarray}
and the electrostatic potential of the ferromagnetic leads is raised by ${\it \Delta} V_{c}$ against the non-magnetic leads. The work function $W$ of the combined system, however, remains constant irrespective of the magnitude of the applied magnetic field. It is because the nonmagnetic lead is connected to the ground, the difference of the electrostatic potential between the nonmagnetic lead and the ground is fixed, and, therefore, the electrochemical potential $\omega$ measured from the vacuum level at infinity is fixed over the whole combined lead independent of the applied magnetic field. The overall change in the electron system of the combined lead caused by the magnetic field is a decrease in the electrostatic-potential energy of electrons in the ferromagnetic leads by $-e{\it \Delta}V_{\rm c}(H)$ (and, of course, the increase in the chemical potential by ${\it \Delta}\zeta(H)$). 
 
The influence of this effect on the SET device is the change in the electrostatic potential of the island electrode against the non-magnetic lead electrodes through its capacitive coupling with the adjoining ferromagnetic leads. For example, in the situation of Fig. 2, the change in the electrostatic potential of the island electrode becomes ${\it \Delta} V_{\rm c}(H)\cdot 2{\cal C}/(2{\cal C}+{\cal C}_{\rm g})$, where ${\cal C}$ is the capacitance of each tunnel junction and ${\cal C}_{\rm g}$ is the capacitance of the gate capacitor.

We should note the mechanism is effective even if all the leads except the gate electrode are ferromagnetic. This is because the spatial variation of the magnetic field causes the gradient of electrostatic potential in a ferromagnetic lead. The electrostatic potential of the parts of the lead electrodes, which are adjacent to the island electrode via the tunnel barriers, against the ground or against the gate electrode is determined by the magnetic field at the junctions irrespective of the kind of metals in between. If the gate electrode is ferromagnetic as well, the variation of its electrostatic potential due to the magnetic field should be included in the consideration.

\section{The Thermodynamic Potential for the Island Electrode}
It has been indicated that conductance of SET-type devices at small bias voltages ($eV_{\rm bias}\ll k_{\rm B}T$) are predominated by a distribution function, $P_{\rm eq}(N)$, of the electron number in the island electrode in equilibrium. \cite{Beenakker,KulShek} The distribution function is expressed by using a Gibbs thermodynamic potential $\Omega(N)$  of the island electrode as \cite{Beenakker,KulShek} 
\begin{equation}
P_{\rm eq}(N)=
\frac{
{\rm exp}[-\Omega(N)/k_{\rm B}T]
}
{
\mathop{{\sum}}_{N=0}^{\infty}{\rm exp}[-\Omega(N)/k_{\rm B}T]
}.
\end{equation}
Now we consider the Gibbs thermodynamic potential of the island electrode of the ferromagnetic SET in a magnetic field. For the sake of clarity, we consider especially the case of ferromagnetic SET in Fig. 2; the system is composed of an island electrode of ferromagnetic metal A, adjoining lead electrodes of ferromagnetic metal B, other electric circuit made of non-magnetic metal C and a source of electrostatic-potential difference for the gate. We assume the circuit is connected to the ground at the non-magnetic lead C and is placed in a space with no background gradient of the electrostatic potential. 

First we examine the electrochemical potentials of the majority- and the minority-spin conduction electrons in the island electrode A. The chemical potentials of the majority- and the minority-spin electrons, $\zeta_{\pm}(H)$, measured from the bottom of each conduction band, are expressed as
\begin{equation}
\zeta_{\pm}(H)=\zeta(H)\pm{\it \Delta}\pm\delta(H).
\end{equation}
If the shift of the electron bands relative to the vacuum level due to the charging effect is not considered, the energies of the bottoms of the conduction subbands, ${\cal E}_{\pm}$, measured from the vacuum level, are expressed as
\begin{equation}
{\cal E}_{\pm}=-{\cal E}_{c}^{0}\mp{\it \Delta}\mp\delta(H),
\end{equation}
where ${\cal E}_{c}^{0}$ means the energy distance between the vacuum level and the midpoint between the bottoms of the conduction bands without the charging effect. From eqs. (4.2) and (4.3), the electrochemical potentials of the majority- and the minority-spin electrons, $\omega_{\pm}(H)$, measured from the vacuum level, become
\begin{eqnarray}
\omega_{\pm}(H)&=\zeta_{\pm}(H)+{\cal E}_{\pm} \nonumber \\
&=\zeta(H)-{\cal E}_{c}^{0} \nonumber \\
&\equiv\omega_{\rm A}(H).
\end{eqnarray}
The electrochemical potential $\omega_{\rm A}(H)$ is equivalent to $-W_{\rm A}(H)$, where $W_{\rm A}(H)$ is the work function of the ferromagnet A in magnetic field $H$. Therefore Gibbs free energy of the conduction electrons without consideration of the charging effect becomes 
\begin{equation}
N_{+}\omega_{+}(H)+N_{-}\omega_{-}(H)=N\omega_{\rm A}(H).
\end{equation}
It should be noted here that now energy of an electron is measured from the vacuum level at infinity.

Next we take the charging effect into account. If the self capacitance ${\cal C}_{\Sigma}$ of the island electrode is small, a change in the electron number from that of the electrostatic neutrality causes a considerable change in the electrostatic potential of the island electrode relative to the lead electrodes and electrostatic energy is accumulated in the system. Electrostatic-potential energy of conduction electrons is also changeable homogeneously by external sources. We label the electrostatic energy of the island as $U$, which includes both components due to the charging effect and external sources. In the case of Fig. 2 the external source is the electrostatic-potential-difference source for the gate, \cite{OSC} and $U$ is a function of $N$, $V_{\rm g}$ and $H$. Then the Gibbs free energy of the island electrode becomes $N\omega_{\rm A}(H)+U(N, V_{\rm g}, H)$, and the Gibbs thermodynamic potential, $\Omega(N, V_{\rm g}, H)$, in this case is expressed as
\begin{equation}
\Omega(N, V_{\rm g}, H)=N\omega_{\rm A}(H)+U(N, V_{\rm g}, H)-N\omega_{\rm BC},
\end{equation}
where $\omega_{\rm B\rm C}$ denotes the electrochemical potential of the combined leads B and C measured from the vacuum level. Magnitude of $\omega_{\rm B\rm C}$ is equivalent with the work function of the combined system B and C, $W_{\rm BC}$, which is independent of the applied field, as discussed in $\S$3.

In the present discussion, we assumed $\omega_{\rm A}$ ($\omega_{\pm}$) has no $N$ dependence, which is reasonable because, although the metallic island electrode is of the mesoscopic size, it still contains a large number of electrons. It results in a dense density of states at the Fermi energy. For example, in Ni of dimensions of 100nm$\times$2$\mu$m$\times$20nm, the density of states at the Fermi energy is about 9$\times 10^{9}$ states Ryd${}^{-1}$ (see Table I). Thus a change in $N$ of unity causes a change in the chemical potential of the order of 10${}^{-9}$ eV ($\sim$ 10${}^{-5}$K), which is negligibly small compared with the magnitude of a typical charging energy ($\sim$ 1K).
 
For the sake of clarity and completeness of the discussions, we hereafter consider the polarity of the magnetic field explicitly, and introduce spatial polarity into the spin polarization of the electronic densities of states. We take the direction of positive magnetic field as {\it z} direction and define a spin polarization with spatial polarity, ${\cal P}'$, as
\begin{equation}
{\cal P}'=\left\{
\begin{array}{cc}
+{\cal P}  & {\rm (\, magnetization \ \ is \ \ parallel \ \ to \ \ {\it z}\, )}\\
-{\cal P}  & {\rm (\, magnetization \ \ is \ \ antiparallel \ \ to \ \ {\it z}\, )}.\\
\end{array}
\right.
\end{equation}
According to this notation, ${\cal P}'H$ becomes ${\cal P}|H|$ if the magnetization and the applied magnetic field is parallel and becomes $-{\cal P}|H|$ if they are antiparallel irrespective of whether $H$ is positive or negative. We call ${\cal P}'$ simply as the 'spin polarization' hereafter.

Using this notation, $\omega_{\rm A}(H)$ is expressed, according to the consideration in $\S$2, as follows:
\begin{equation}
\omega_{\rm A}(H)=\omega_{\rm A}(0)-{\cal P}'_{\rm A}g_{\rm A}\mu_{\rm B}H/2,
\end{equation}
where ${\cal P}'_{\rm A}$ denotes the spin polarization of the ferromagnet A in zero magnetic field. \cite{omegaA}
The electrostatic energy $U(N, V_{\rm g}, H)$ is expressed as follows:
\begin{equation}
U(N, V_{\rm g}, H)=\frac{(N-N_{0})^{2}e^{2}}{2{\cal C}_{\Sigma}}-Ne\phi_{\rm ext}(V_{\rm g}, H),
\end{equation}
where $N_{0}$ denotes the number of electrons when the island electrode is electrostatically neutral, ${\cal C}_{\Sigma}=2{\cal C}+{\cal C}_{\rm g}$ and 
\begin{equation}
\phi_{\rm ext}(V_{\rm g}, H)=\frac{{\cal C}_{\rm g}}{{\cal C}_{\Sigma}}V_{\rm g}+\frac{2{\cal C}}{{\cal C}_{\Sigma}}V_{\rm c}(H).
\end{equation}
Here $V_{\rm g}$ is the electrostatic potential of the gate electrode relative to the non-magnetic lead C due to the electrostatic-potential-difference source for the gate and $V_{\rm c}(H)$ denotes the contact potential of the ferromagnetic leads B relative to the electrostatic potential of the non-magnetic lead C in magnetic field $H$. According to the discussion in $\S$3, $V_{\rm c}(H)$ is expressed as follows:
\begin{equation}
V_{\rm c}(H)=V_{\rm c}(0)-{\cal P}'_{\rm B}g_{\rm B}\mu_{\rm B}H/2e,
\end{equation}
where $V_{\rm c}(0)$ is the contact potential of the ferromagnet B against the non-magnetic metal C in zero magnetic field, and ${\cal P}'_{\rm B}$ denotes the spin polarization of the ferromagnet B in zero magnetic field.

After rearrangement of all the entities in $\Omega(N, V_{\rm g}, H)$, we obtain the next expression:
\begin{eqnarray}
\Omega(N, V_{\rm g}, H)=\frac{e^{2}}{2{\cal C}_{\Sigma}}
{
\left[
N-\frac{{\cal C}_{\Sigma}}{e}
\left\{
\left(\frac{{\cal C}_{\rm g}}{{\cal C}_{\Sigma}}\right)V_{\rm g}
-\left(\frac{2{\cal C}}{{\cal C}_{\Sigma}}\right)
{\cal P}'_{\rm B}g_{\rm B}\mu_{\rm B}H/2e
+{\cal P}'_{\rm A}g_{\rm A}\mu_{\rm B}H/2e\right.\right.} \nonumber \\
\left.\left.
+N_{0}e/{\cal C}_{\Sigma}
+\omega_{\rm B\rm C}/e
-\omega_{\rm A}(0)/e
+\left(\frac{2{\cal C}}{{\cal C}_{\Sigma}}\right)V_{\rm c}(0)
\frac{}{}
\right\}
\right]^{2}
+(\ {\rm terms \ independent \ of} \ N\ ).
\end{eqnarray}

\section{Magneto-Coulomb Oscillation in the Ferromagnetic SET}
\subsection{The period of the magneto-Coulomb oscillation}
According to the theory of the Coulomb oscillation, \cite{Beenakker,KulShek} conductance of an SET in the linear response regime is expressed as follows:
\begin{equation}
G=\frac{1}{k_{\rm B}T}\frac{1}{R_{\rm T}^{l}+R_{\rm T}^{r}}\mathop{\sum}_{N=0}^{\infty}
P_{\rm eq}(N, V_{\rm g}, H)h({\it \Delta}\Omega),
\end{equation}
with
\begin{equation}
{\it \Delta}\Omega=\Omega(N, V_{\rm g}, H)-\Omega(N-1, V_{\rm g}, H),
\end{equation}
where $R_{\rm T}^{l,r}$ mean the tunnel resistances of the left and the right tunnel junctions and $P_{\rm eq}(N, V_{g}, H)$ denotes the statistical probability in equilibrium of $N$ electrons being in the island electrode in magnetic field $H$, and is expressed as
\begin{equation}
P_{\rm eq}(N, V_{\rm g}, H)=
\frac{
{\rm exp}[-\Omega(N, V_{\rm g}, H)/k_{\rm B}T]
}
{
\mathop{{\sum}}_{N=0}^{\infty}{\rm exp}[-\Omega(N, V_{\rm g}, H)/k_{\rm B}T]
}.
\end{equation}
The function $h$ has the form as
\begin{equation}
h(x)=\frac{x}{1+{\rm exp}(x/k_{\rm B}T)}.
\end{equation}
One can see from the expression of $\Omega(N, V_{\rm g}, H)$ in eq. (4.12) that, if the value in the square brackets in eq. (4.12) is equivalent by a modulus of unity, the sum in eq. (5.1) and, therefore, conductance of the SET become equivalent. This fact indicates the periodic nature of the conductance of the SET against $V_{\rm g}$ or $H$. Thus, as a condition which determine the period of the magneto-Coulomb oscillation, ${\it \Delta} H$, we obtain the next equation:
\begin{equation}
\frac{{\cal C}_{\Sigma}}{e}
\left\{
{\cal P}'_{\rm A}g_{\rm A}\mu_{\rm B}{\it \Delta} H/2e
-\left(\frac{2{\cal C}}{{\cal C}_{\Sigma}}\right)
{\cal P}'_{\rm B}g_{\rm B}\mu_{\rm B}{\it \Delta} H/2e
\right\}=1.
\end{equation}
As a result, the period of the magneto-Coulomb oscillation is expressed as follows:
\begin{equation}
{\it \Delta} H=\left(\frac{e^{2}}{\mu_{\rm B}}\right)
\frac{2}{{\cal C}_{\Sigma}{\cal P}'_{\rm A}g_{\rm A}
-2{\cal C} {\cal P}'_{\rm B}g_{\rm B}}.
\end{equation}

\subsection{The equi-phase lines in the $V_{\rm g}$-$H$ diagram}
It is also evident from the expression of $\Omega(N, V_{\rm g}, H)$ and eqs. (5.1)--(5.3) that, if the value in the square brackets in eq. (4.12) is kept constant with fixed $N$ by tuning both $V_{\rm g}$ and $H$, the conductance of the SET is kept constant. This tuning is realized by the condition such as
\begin{equation}
\left(\frac{{\cal C}_{\rm g}}{{\cal C}_{\Sigma}}\right)V_{\rm g}
-\left(\frac{2{\cal C}}{{\cal C}_{\Sigma}}\right)
{\cal P}'_{\rm B}g_{\rm B}\mu_{\rm B}H/2e
+{\cal P}'_{\rm A}g_{\rm A}\mu_{\rm B}H/2e={\rm constant}.
\end{equation}
This equation defines the equi-phase lines in the $V_{\rm g}$-$H$ diagram. Namely they are expressed as
\begin{equation}
V_{\rm g}=
\left(\frac{\mu_{\rm B}}{e}\right)
\frac{2{\cal C} {\cal P}'_{\rm B}g_{\rm B}
-{\cal C}_{\Sigma}{\cal P}'_{\rm A}g_{\rm A}}{2{\cal C}_{\rm g}}
H
+ {\rm constant}.
\end{equation}

\subsection{Expected $V_{\rm g}$-$H$ diagrams in several ferromagnetic SET's}
Here we discuss some expected $V_{\rm g}$-$H$ diagrams of ferromagnetic SET's based on the discussions in the previous subsections. First we consider the case of Fig. 3(a): the SET is a non-magnetic metal/ferromagnetic metal/non-magnetic metal (N/F/N) SET. We assume here the polarization ${\cal P}$ is negative in the ferromagnet. From eq. (5.6) the period of the magneto-Coulomb oscillation is ${\it \Delta} H=(2e^{2}/\mu_{\rm B})/({\cal C}_{\Sigma}|{\cal P}|g)$. The equi-phase lines are expressed as
\[
\makebox[-3.7cm]{}
V_{\rm g}
=-\left(\frac{\mu_{\rm B}}{e}\right)\left(\frac{{\cal C}_{\Sigma}}{2{\cal C}_{\rm g}}\right)g{\cal P}'H
+ {\rm constant}
\]
\begin{equation}
\makebox[3cm]{}=\left\{
\begin{array}{l}
\left(\frac{\mu_{\rm B}}{e}\right)\left(\frac{{\cal C}_{\Sigma}}{2{\cal C}_{\rm g}}\right)g|{\cal P}||H|
+ {\rm constant} \nonumber \\
\makebox[3cm]{}(\,{\rm if \ \ magnetization \ \ and \ \ }H{\rm \ \ are \ \  parallel}\,) \nonumber \\
-\left(\frac{\mu_{\rm B}}{e}\right)\left(\frac{{\cal C}_{\Sigma}}{2{\cal C}_{\rm g}}\right)g|{\cal P}||H|
+ {\rm constant} \nonumber \\
\makebox[3cm]{}(\,{\rm if \ \ magnetization \ \ and \ \ }H{\rm \ \ are \ \  antiparallel}\,).
\end{array}
\right.
\end{equation}
For example, suppose one is sweeping a magnetic field from positive strong magnetic field to negative one. In positive strong magnetic fields, the direction of the magnetization is parallel to the applied magnetic field, ${\cal P}'=+{\cal P}$ and the equi-phase lines in this region is expressed by the upper relation of eq. (5.9). Once the field is reversed, the magnetization and the field becomes antiparallel as long as the magnitude of the field is smaller than the coercive force, $H_{\rm c}$, of the ferromagnet. In this region, ${\cal P}'=-{\cal P}$ and the equi-phase lines are expressed by the lower relation of eq. (5.9). At $H=-H_{\rm c}$, the direction of the magnetization changes parallel to the applied field, there occurs an instantaneous change in the value of the expression in the braces in eq. (4.12) by $2g{\cal P}\mu_{\rm B}H_{\rm c}/e$, a phase jump occurs in the $V_{\rm g}$-$H$ diagram, now ${\cal P}'=+{\cal P}$ and the equi-phase lines are once again expressed by the upper relation in eq. (5.9). The expected equi-phase lines for the peaks (solid lines) and valleys (broken lines) of the conductance are schematically depicted in Fig. 3(b).
\begin{figure}
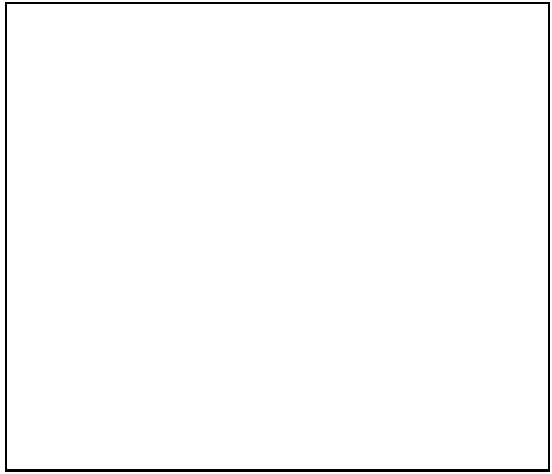

\figureheight{6cm}
\caption{Expected $V_{\rm g}$-$H$ diagrams of two kinds of ferromagnetic SET's. The polarization ${\cal P}$ of the ferromagnet is assumed to be negative. The solid lines are the peaks of the conductance of the SET and broken lines are its valleys. $H_{\rm c}$ indicates the coercive force of the ferromagnet. ${\it \Delta} V_{\rm g}$ is the period of the Coulomb oscillation. ${\it \Delta} H$ and ${\it \Delta} H'$ are the periods of the magneto-Coulomb oscillations in ferromagnetic SET's of the type (a) and (c), respectively.}
\label{figure:3}
\end{figure}

Next we examine the case of an F/N/F ferromagnetic SET such as in Fig. 3(c). Here we once again assume ${\cal P}$ is negative in the ferromagnets. In this case, from eq. (5.6), the period of the magneto-Coulomb oscillation, ${\it \Delta} H'$, becomes ${\it \Delta} H'=(e^{2}/\mu_{\rm B})/({\cal C} |{\cal P}|g)$, and the equi-phase lines in the $V_{\rm g}$-$H$ diagram are expressed as
\[
\makebox[-3.7cm]{}
V_{\rm g}
=\left(\frac{\mu_{\rm B}}{e}\right)\left(\frac{{\cal C}}{{\cal C}_{\rm g}}\right)g{\cal P}'H
+ {\rm constant}
\]
\begin{equation}
\makebox[3cm]{}=\left\{
\begin{array}{l}
-\left(\frac{\mu_{\rm B}}{e}\right)\left(\frac{{\cal C}}{{\cal C}_{\rm g}}\right)g|{\cal P}||H|
+ {\rm constant} \nonumber \\
\makebox[3cm]{}(\,{\rm if \ \ magnetization \ \ and \ \ }H{\rm \ \ are \ \  parallel}\,) \nonumber \\
\left(\frac{\mu_{\rm B}}{e}\right)\left(\frac{{\cal C}}{{\cal C}_{\rm g}}\right)g|{\cal P}||H|
+ {\rm constant} \nonumber \\
\makebox[3cm]{}(\,{\rm if \ \ magnetization \ \ and \ \ }H{\rm \ \ are \ \  antiparallel}\,).
\end{array}
\right.
\end{equation}
In this case the slope of the lines is in opposite polarity to the case of the N/F/N ferromagnetic SET. The expected equi-phase lines in this case for the peaks (solid lines) and valleys (broken lines) of the conductance are schematically depicted in Fig. 3(d). 

So far we have assumed that the ferromagnets under consideration have single domains. However, in a real ferromagnet of not extremely small size, there exists domain structure in general. In this case the magnetization reversal may occur in a finite range of magnetic field, or/and it may occur in successive discrete magnetic fields. Partial reversal of magnetization near the junction electrode due to the change in the domain structure as well affects the operation of the SET through the same mechanism.
First, we consider an island electrode that is composed of many magnetic domains. We specify the magnetic domain by the number $i$, name magnetization of the $i$-th domain {\boldmath $m$}$_{i}$, its self-capacitance ${\cal C}_{i}$ in a magnetic field {\boldmath $H$}. \cite{vector} The shift of chemical potential of the $i$-th domain, ${\it \Delta}\zeta_{i}$, in {\boldmath $H$} is ${\it \Delta}\zeta_{i}=-{\cal P}g\mu_{\rm B}\,\frac{\mbox{\boldmath $m$}_{i}\cdot\mbox{\boldmath $H$}}{2\, |\mbox{\boldmath $m$}_{i}|}$ as considered in $\S$2. We assume that, after transfer of electrons among the domains to equilibrate their electrochemical potentials in {\boldmath $H$}, excess charge, $q_{i}$, is accumulated in the $i$-th domain. Then the shift of the electrochemical potential of the $i$-th domain, ${\it \Delta}\omega_{i}$, from that in zero magnetic field is obtained as ${\it \Delta}\omega_{i}={\it \Delta}\zeta_{i}-e\, q_{i}/{\cal C}_{i}$. This quantity should be the same among all the domains in equilibrium and represents the shift of the electrochemical potential of the island electrode, ${\it \Delta}\omega$, in equilibrium. By using a condition, $\sum_{i}^{N} q_{i}=0$, with $N$ being the total number of magnetic domains in the island electrode in {\boldmath $H$}, and noting $\sum_{i}{\cal C}_{i}={\cal C}_{\Sigma}$, we obtain
\begin{equation}
{\it \Delta}\omega=-\frac{{\cal P}g\mu_{\rm B}}{2{\cal C}_{\Sigma}}\sum_{i}^{N}\frac{{\cal C}_{i}\,\mbox{\boldmath $m$}_{i}\cdot \mbox{\boldmath $H$}}{|\mbox{\boldmath $m$}_{i}|}.
\end{equation}
This means that each domain affects the shift of the electrochemical potential by the factor of ${\cal C}_{i}/{\cal C}_{\Sigma}$. In real devices, we usually observe that capacitances of the junctions are much larger than the other contributions to ${\cal C}_{\Sigma}$. In such cases, we can conclude, from the above expression, that the shift of the electrochemical potential of the island electrode is almost determined by the magnetizations around the junction electrodes because self-capacitances of domains around the junction electrode is much larger than the others through the electrostatic coupling with the adjoining counter electrodes.

Next we examine multi-domain lead electrodes. Here again we specify the domains by the number $i$, name magnetization of the $i$-th domain {\boldmath $m$}${}_{i}$ and the mutual capacitance between it and the island electrode ${\cal C}'_{i}$. In this case, as discussed in $\S$3, the electrochemical potential of each domain is unchangeable by application of magnetic field. The shift of the electrostatic potential of the $i$-th domain, ${\it \Delta}V_{i}$, is expressed as ${\it \Delta}V_{i}=-{\cal P}g\mu_{\rm B}\,\frac{\mbox{\boldmath $m$}_{i}\cdot \mbox{\boldmath $H$}}{2e\, |\mbox{\boldmath $m$}_{i}|}$. The change in the electrostatic potential of the island electrode, ${\it \Delta}V$, due to ${\it \Delta}V_{i}$'s is expressed as
\begin{equation}
{\it \Delta}V=-\frac{{\cal P}g\mu_{\rm B}}{2e\,{\cal C}_{\Sigma}}\sum_{i}^{N'}\frac{{\cal C}'_{i}\,\mbox{\boldmath $m$}_{i}\cdot \mbox{\boldmath $H$}}{|\mbox{\boldmath $m$}_{i}|},
\end{equation}
where $N'$ is the number of the domains in the ferromagnetic lead electrodes in {\boldmath $H$}. This expression indicates that changes in magnetization around the junction electrodes are most effective in changing the electrostatic potential of the island electrode through the large mutual capacitance between the island electrode and the domains around the junction electrode.
Thus changes in the domain structures in the ferromagnetic island and lead electrodes around the junction electrodes could cause sequential phase shifts in the $V_{\rm g}$-$H$ diagram around their coercive forces in sweeping the magnetic field.

On the other hand, if the ferromagnetic component of the ferromagnetic SET is a single crystal, the period of the magneto-Coulomb oscillation and the slope of the equi-phase lines in the $V_{\rm g}$-$H$ diagram would show non-monotonic dependences on the magnetic field, which have oscillatory structures, in high magnetic fields. It is because the electrochemical potential of the conduction electrons in a metallic single crystal shows an oscillatory dependence on the magnetic field, reflecting the Landau quantization of the electron orbits in high magnetic fields. \cite{Gold,LifKos} Such oscillatory electrochemical potential has been actually observed in contact potential measurements. \cite{Gold} However, in a polycrystalline metal such as an evaporated thin film, there exist crystal domains of various orientations, and such oscillatory dependence of the electrochemical potential should be averaged out. Such effect may be observed in a SET of its island electrode made of a single fine particle.

\section{Consideration on the Chemical-Potential Shift}
As described in the previous sections, magnetic-field-induced shifts of the chemical potentials in the ferromagnetic components of the ferromagnetic SET cause the magneto-Coulomb oscillation. So far we have considered the chemical-potential shift in a ferromagnet only through the bare spin-Zeeman effect. Actually there are contributions to the shift due to other effects. Thus the polarization ${\cal P}$ in the above sections should be replaced by some effective one, ${\cal P}_{\rm eff}$. In this section we discuss more realistic estimation of the magnitude of the chemical-potential shift in magnetic fields. Because numerical calculations based on the one-electron approximation have made a considerable success in describing the electronic properties of transition metals, \cite{Shimizu} we here discuss it in the one-electron approximation.

\subsection{Exchange-correlation correction}
In \S2-5, we considered that the bare applied magnetic field alone affects the electron system as a perturbation. However, in 3$d$ transition metals, the effective one-electron potential also depends on the magnetic field and, as a result, the perturbation is enhanced. Here we examine this effect from the point of view of the local spin density approximation (LSDA). As described in $\S$2, population of electrons in the states near Fermi energy changes in magnetic fields. As a result, the local electron density, $n(\mbox{\boldmath$r$})$, and the local spin density, $m(\mbox{\boldmath$r$})$, change. In the one-electron approximation, an electron is assumed to be situated in an effective potential composed of contribution by the positive ions and that by the other electrons. Therefore, changes in $n$ and $m$ would make the effective potential which an electron feels change. In LSDA \cite{Yamada} such a change in the effective one-electron potential is described by variations in the Hartree potential, $v_{\rm H}(\mbox{\boldmath$r$})$, and the exchange-correlation potential, $v_{\rm xc}^{\sigma}(\mbox{\boldmath$r$})$ ($\sigma$ specifying the spin), which are functionals of $n$ and $m$. Namely, an electron feels extra perturbation of ${\it \Delta}v_{\rm H}(\mbox{\boldmath$r$})+{\it \Delta}v_{\rm xc}^{\sigma}(\mbox{\boldmath$r$})$ in magnetic fields. The so-called Stoner enhancement of the magnetic susceptibility of the transition metals is considered to be caused by this extra perturbation. \cite{Yamada}

According to the homogeneous enhancement model, which is so far the most successful method to calculate the high-field spin susceptibility of the ferromagnetic metals, \cite{Yamada,Yasui} it is concluded that the induced change in $n$ vanishes although that in $m$ is enhanced in homogeneous magnetic fields. This causes an correction to the change in the chemical potential through the variation in $v_{\rm xc}^{\sigma}(\mbox{\boldmath$r$})$. We obtain the next expression for the change in the chemical potential due to the spin, ${\it \Delta}\zeta_{\,\rm s}$, in this approximation:
\begin{equation}
{\it \Delta}\zeta_{\,\rm s}(H)=-{\cal P}g\mu_{\rm B}H(1+{\it \Upsilon})/2,
\end{equation}
where
\begin{equation}
{\it \Upsilon}=-\left(\frac{1}{{\cal P}^{2}}-1\right)(\alpha_{3}+\alpha_{4}) \chi_{\rm hf}.
\end{equation}
In the above expression, $\chi_{\rm hf}$ and $\alpha_{3,4}$ respectively represent the high-field spin susceptibility and two of the four contributions to the molecular-field coefficient calculated in this approximation in ref. 19. Details of the derivation of the above result are presented in Appendix A. 

\subsection{Contribution of the orbital magnetic moment}
Orbital magnetic moments of conduction electrons also contribute to the shift of the chemical potential by the change in their magnetic energies in magnetic fields. As it is suggested in the calculation of the high-field magnetic susceptibility that the effect of the exchange-correlation correction is negligible, \cite{Yasui} we calculate the orbital contribution to the chemical-potential shift, ${\it \Delta}\zeta_{\rm orb}$, only by the bare perturbation of the external magnetic field. According to the derivation in Appendix B, ${\it \Delta}\zeta_{\rm orb}$ up to the 1st order in $H$ is given by
\begin{equation}
{\it \Delta}\zeta_{\rm orb}=\frac{\mu_{\rm B}H}{\rho_{+}+\rho_{-}}
\sum_{i\sigma}\frac{{\rm d}f({\epsilon}_{i\sigma}-\zeta_{0})}{{\rm d}{\epsilon}_{i\sigma}}<i\sigma|{L}_{z}|i\sigma>,
\end{equation}
where $i$ specifies the one-electron eigen state and $\zeta_{0}=\zeta(H={\rm 0})$. As can be seen in the above expression, orbital magnetic moments belonging to the states around the Fermi energy cause the shift of the chemical potential. Average of them is not necessarily zero in real metals because of the spin-orbit coupling.

Taking into account the results in this and the above subsections, ${\cal P}_{\rm eff}$ at low temperatures could be expressed as
\begin{equation}
{\cal P}_{\rm eff}={\cal P}(\zeta_{0})(1+{\it \Upsilon})
+(2/g)<\, L_{z}\,>_{0}.
\end{equation}
Here $<\, L_{z}\,>_{0}$ indicates the average of the eigen values of $L_{z}$ of the electron states at the Fermi level in zero magnetic field.

\section{ Experimental Observation of the Magneto-Coulomb Oscillation in a Ni/Co/Ni SET }
In this section we present experimentally observed results of the magneto-Coulomb oscillation in the Ni/Co/Ni SET showing hysteretic behavior against the sweep direction of the magnetic field. The device was the same as that used in ref. 4. It is composed of Ni lead electrodes, NiO tunnel barriers and a Co island electrode fabricated on a Si wafer with a Ag gate electrode on the rear surface of the wafer. The AFM micrograph of the device is presented in Fig. 4. The area of the junctions is about 0.02$\mu$m${}^{2}$, resistance of the junction, $R_{\rm T}$, is about 35k$\Omega$ and the charging energy, $e^{2}/2{\cal C}_{\Sigma}$, is estimated to be 25-50$\mu$eV. As seen in the figure, there exist many small electrodes of the similar shapes with the device itself. However, they are not connected to the measurement pads and, therefore, do not contribute to the electric conduction. Further details on the sample are described in ref. 4. 
\begin{figure}
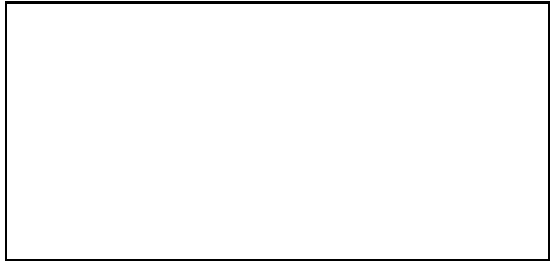

\figureheight{3.2cm}
\caption{An AFM micrograph of the Ni/Co/Ni SET device. The electrodes surrounded by a broken-line square compose the device. The other small electrodes are dangling and do not contribute to the electric conduction. The black spots on both sides of the Co electrode are the tunnel junctions. The gate electrode made of Ag is on the rear surface of the chip.}
\label{figure:4}
\end{figure}

Measurements were made by sweeping the applied magnetic field slowly and scanning the applied gate voltage quickly with the excitation current of 20pA at 12.5Hz to measure the zero-bias resistance of the device in a dilution refrigerator at 20mK. The direction of the applied magnetic field was parallel to the long axis of the electrodes.

In Figs. 5(a) and 5(b) are presented the observed results, which show the zero-bias resistance of the device by the gray-scale plot on the $V_{\rm g}$-$H$ diagrams, for both cases of sweeping $H$ to ($\pm$) directions. (The result in sweeping $H$ to ($-$) direction is the same one as in ref. 4.) In both cases of sweep directions, the Magneto-Coulomb oscillation is observed clearly. Near zero magnetic field, particularly within $\pm$10kOe, there exist many jumps in the equi-phase lines. Moreover, the position of these jumps seems to show hysteretic behavior against the sweep direction of the magnetic field. Further, there exist solitary phase jumps at high magnetic fields. The magnetic fields at which these high-field phase jumps occur are not systematic. They changed sweep by sweep.
\begin{figure}
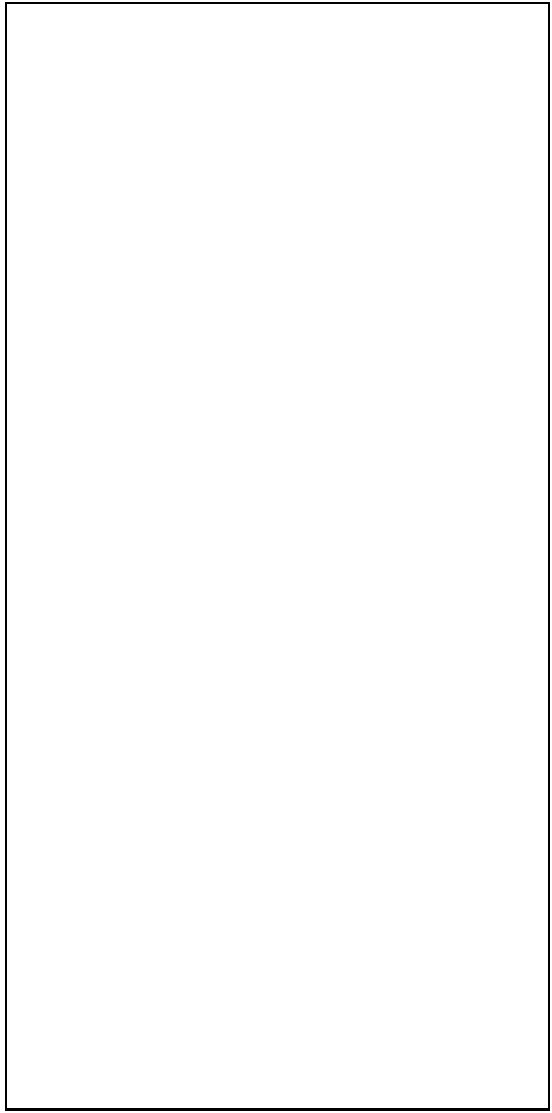

\figureheight{14.5cm}
\caption{Hysteretic behavior of the magneto-Coulomb oscillation in a Ni/Co/Ni SET against the sweep direction of the magnetic field. Bright (dark) regions correspond to high (low) resistance. The arrow above each frame shows the sweep direction of the magnetic field.}
\label{figure:5}
\end{figure}

\section{ Interpretation of the Experimental Results}
Now we consider the case of the experiment in \S7 and in ref. 4 in detail. The key quantity in the mechanism discussed above is the spin polarization ${\cal P}$ of the densities of states. However, ${\cal P}$ is not a directly measurable quantity, and we have to rely on the band calculation. There have been many band calculations on Ni and also several on Co. They have following common features: the Fermi energy of the electrons lies just above the majority-spin $d$ bands in both materials, in the middle of the minority-spin $d$ bands in Co and near the upper-most of the minority-spin $d$ bands in Ni with a much larger density of states ($\rho_{-}$) than in Co. Thus it is expected that the polarizations are negative in both materials and $|{\cal P}_{\rm Ni}|>|{\cal P}_{\rm Co}|$. Therefore the expected $V_{\rm g}$-$H$ diagram is of the type of Fig. 3(d). This is consistent with the experimental observation.

Some of the band calculations give the densities of states of the majority- and the minority-spin electrons at the Fermi energy. Using these values, we calculated ${\cal P}$'s in both Co and Ni. The results are tabulated in Table I with the values of the densities of states at the Fermi energy per atom ($\rho_{\pm}{}^{({\rm 0})}$). In the case of Co the calculated ${\cal P}$ scatters a little. 
\begin{table}
\caption{Densities of states per atom of the majority- and the minority-spin electrons at the Fermi energies and the polarizations of Co and Ni.}
\label{table:1}
\begin{tabular}{@{\hspace{\tabcolsep}\extracolsep{\fill}}cccc} \hline
material & $\rho_{+}{}^{({\rm 0})*}$  & $\rho_{-}{}^{({\rm 0})*}$ & ${\cal P}$\\ \hline
Co       & 6.64${}^{a}$	       & 16.50${}^{a}$      & $-0.43$\\
(hcp)    & 2.3${}^{b}$         & 9.0${}^{b}$        & $-0.59$\\
         & 2.29${}^{c}$        & 10.36${}^{c}$      & $-0.64$\\ \hline
Ni	 & 2.8${}^{d}$	       & 22.3${}^{d}$       & $-0.78$\\
(fcc)    & 2.3${}^{b}$         & 22.4${}^{b}$       & $-0.81$\\
         & 2.860${}^{e}$       & 25.585${}^{e}$     & $-0.80$\\ \hline
\end{tabular}
${}^{*}$ in a unit of states ${\rm Ryd}^{-1}{\rm atom}^{-1}$.\\
a) from ref. 21, b) from ref. 22, c) from ref. 23, d) from ref. 24, e) from ref. 25.
\end{table}

On the other hand, sufficient information is not available on the enhancement factor ${\it \Upsilon}$ defined in $\S$6 and on the orbital contribution. According to the numerical calculations in refs. 19 and 20, the factor $(\alpha_{3}+\alpha_{4})\chi_{\rm hf}$ in eq. (6.2) is $-$0.16 for Fe and roughly $-$0.3 for Ni, which gives rather small effect on ${\it \Delta}\zeta_{\,\rm s}$. The same conclusion is expected for Co. So we will neglect it in the following numerical estimation. The orbital magnetic moment in metals appears due to the spin-orbit interaction. However, the magnitude of the spin-orbit interaction in the 3{\it d} transition metals is not very large; actually, the orbital magnetic moments in these metals are very small. \cite{FrNoEc,EbStGy,KruSpr} So we will neglect the orbital contribution as well.

The period of the magneto-Coulomb oscillation and the slope of the equi-phase lines in the $V_{\rm g}$-$H$ diagram depend also on the parameters in the circuit, namely, the capacitances of the tunnel junctions and that of the gate capacitor. In our device, capacitance of the gate capacitor ${\cal C}_{\rm g}$ is estimated to be about 0.60 aF from the period of the Coulomb oscillation. On the other hand, the capacitances of the tunnel junctions are not determined accurately in the experiment, because the tunnel resistance of the junctions are rather low and smearing of the current ($I$)-voltage ($V$) characteristic occurs. \cite{Ono2} We assumed that the capacitances of both tunnel junctions were equivalent and estimated the magnitude of the sum of the capacitances $2{\cal C}$ as 1.6-3.2 fF from the offset voltage in the $I$-$V$ characteristic of the device. Thus we calculate the period of the magneto-Coulomb oscillation from the next equation:
\begin{equation}
{\it \Delta} H=\left(\frac{e^{2}}{g\mu_{\rm B}}\right)\frac{1}{{\cal C}({\cal P}_{\rm Co}-{\cal P}_{\rm Ni})}.
\end{equation}
From Table I, ${\cal P}_{\rm Co}-{\cal P}_{\rm Ni}$ becomes 0.14-0.38. In the experiment, the magnetic field is applied parallel to the long axis of the thin film electrodes, so we can ignore the demagnetization effect. Thus the expected period of the magneto-Coulomb oscillation ${\it \Delta} H$ and the slope ${\rm d}V_{\rm g}/{\rm d}H$ become 2.0-11.2T and 25-130mV/T, respectively. The observed values, around 2.2T and 120-190mV/T for ${\it \Delta} H$ and ${\rm d}V_{\rm g}/{\rm d}H$, show a considerable consistency with the theory.  

As observed in Fig. 5 there exist many distinct jumps of the equi-phase lines in both positive and negative fields within about $\pm$10kOe. Moreover, they appear mostly on the (+) side when $H$ is swept from ($-$) to (+) and vice versa. We also observed in ref. 4 a hysteretic tunneling magneto-resistance (MR) near $H$=0Oe: in sweeping $H$ from ($-$) to (+), for example, the resistance gradually increases from about $H$=$-$1kOe till $H$=$+1$kOe where the resistance sharply falls. 
This latter phenomenon is ascribed to the so-called {\it magnetic-valve effect}: \cite{Ono1,Ono2,Maekawa,MesTed} because of the difference of the coercive forces of the Co and Ni electrodes, relative orientation of their magnetizations shows magnetic-field dependence. This causes magnetic-field-dependent tunneling probability of electrons between the two electrodes if the electron spins are conserved during the tunneling processes. From the measurement of magnetic induction in ref. 11 of a Ni/NiO/Co tunnel junction, the coercive force of Ni is expected to be smaller than that of Co. Therefore the result on the MR in ref. 4 indicates that, in sweeping $H$ from ($-$) to (+), for example, the magnetization of the Ni electrodes gradually changes its direction from $H=-$1kOe and that of the Co electrode changes sharply at $H$=$+1$kOe in the device. 

Naturally similar hysteretic behavior is expected for the equi-phase lines in the $V_{\rm g}$-$H$ diagram as mentioned in \S\S 5.3. If magnetic field is swept from ($-$) to (+), for example, ${\cal P}'_{\rm Ni}$ changes from $+{\cal P}_{\rm Ni}$ to $-{\cal P}_{\rm Ni}$ gradually around $H$=0kOe and that ${\cal P}'_{\rm Co}$ changes from $+{\cal P}_{\rm Co}$ to $-{\cal P}_{\rm Co}$ rather abruptly around $H$=$+1$kOe. Moreover, as can be seen in Fig. 4, there is a large Ni electrode on the right side adjacent to the narrow Ni electrode. Capacitive coupling between this large Ni electrode and the island electrode may cause extra phase jumps or shifts, which are associated with the change in the domain structure in this electrode, in a wide magnetic field range. 
The jumps of the equi-phase lines in the $V_{\rm g}$-$H$ diagram in the region near $H$=0Oe must be the reflection of these processes of the magnetization reversal in both Ni and Co electrodes.
On the other hand, the solitary phase jumps at high magnetic fields are considered to be ascribed to uncontrollable shifts of the background charges. 

As can be seen in the high field range in Fig. 5, the equi-phase lines in the $V_{\rm g}$-$H$ diagram show a tendency of leveling off deviating from the straight lines. The origin of this nonlinearity is not clear at present. From the semiclassical treatment \cite{Gold,LifKos} of the free energy of a metal in high magnetic field (without considering the exchange-correlation correction), there appears a term of the 2nd order in $H$ in the non-oscillatory shift of the chemical potential:
$
-(g\mu_{\rm B}H)^{2}(\partial\rho/\partial\epsilon)/8\rho,
$
where $\rho=\rho_{+}+\rho_{-}$.
However, based on our self-consistent APW calculation of the density of states, this term is estimated to be too small to explain the above mentioned nonlinearity, although the energy derivative of the density of states at the Fermi energy in Ni is comparatively very large.

\section{Possible Applications}
As described in the previous sections the origin of the ferromagnetic SET can be considered due to the shift of the chemical potential of the ferromagnet in a magnetic field, and its magnitude depends on the electronic properties in the ferromagnet. Taking the discussions in the above sections conversely, we see that the ferromagnetic SET can provide a method of measuring a thermodynamic coefficient $\upsilon$: 
\begin{equation}
\upsilon=\left(\,\frac{\partial \zeta}{\partial H}\,\right)_{N,\,T},
\end{equation}
which is equivalent to another thermodynamic coefficient
$-(\, \partial M / \partial N)_{H,\,T}$
through the thermodynamics. Here $M$ is the magnetization of the ferromagnet. 
This coefficient contains an information on the electronic structure of the ferromagnetic metal complementary to, say, its high-field magnetic susceptibility. 

If one uses a SET device tunnel resistance of whose junctions are large, one can determine the capacitance of the junction accurately from the current-voltage characteristic of the device as well as the capacitance of the gate capacitor, which can be accurately determined from the period of the Coulomb oscillation. Therefore accurate measurement of $\upsilon$ is possible. The measurement procedure is as follows: the appropriate device is a symmetric N/F/N ferromagnetic SET of the type depicted in Fig. 3(a) of rather high tunnel-junction resistance. After applying the SET a strong magnetic field, for example, in $(+)$ direction in order to align the magnetization of the island electrode to the $(+)$ direction, one decreases the applied magnetic field to near zero field and measure the slope ($\partial V_{\rm g}/\partial H$) of the equi-phase line in the $V_{\rm g}$-$H$ diagram. By using the numerical values of the capacitances $C$ and $C_{\rm g}$, which are determined in other measurements, $\upsilon$ is determined from the slope of the equi-phase line by the next relation:
\begin{equation}
\upsilon=e\left(\frac{{\cal C}_{\rm g}}{{\cal C}_{\Sigma}}\right)
\left(\frac{\partial V_{\rm g}}{\partial H}\right)_{H\rightarrow +0}.
\end{equation}

Within the approximation in \S 6, where the exchange-correlation enhancement of the perturbation is included only in the spin part, the expression of $\upsilon$ becomes as
\begin{equation}
\upsilon=-{\cal P}g\mu_{\rm B}\left(1+{\it \Upsilon}\right)/2-\mu_{\rm B}<\,L_{z}\,>_{0}.
\end{equation}
Apparently ${\it \Upsilon}$ contains an information on the electronic structure complementary to the Stoner enhancement factor.
This method would provide a new tool to investigate electronic structure in ferromagnets complementary to the methods such as the field emission, the spin-resolved photoemission, the magneto-optic Kerr-effect, the spin-polarized tunneling \cite{FeuCut} and the high-field susceptibility measurement. In the present stage of experimental technology, fabrication of N/F/N ferromagnetic SET's of variety of ferromagnets with appropriate device parameters is not difficult. Actually, as a first trial, we have made an Al/Co/Al SET with the tunnel barrier of Al${}_{2}$O${}_{3}$ and estimated the magnitude of $\upsilon$ of Co from the measurement in the magnetic-field range where Al is normal-conducting. \cite{Ono3}

\section{Conclusion}
We have shown that magnetic-field-induced variations in the chemical potentials in ferromagnetic components of the ferromagnetic SET changes the free energy of its island electrode and cause a conductance oscillation (magneto-Coulomb oscillation) similar to the Coulomb oscillation by sweeping an applied magnetic field. The recently observed magneto-Coulomb oscillation in a Ni/Co/Ni ferromagnetic SET can be consistently interpreted in this mechanism. 
Some related phenomena, 
magnitude of the magnetic-field-induced variation of the chemical potential of the conduction electrons in a ferromagnetic metal, 
and an application of the ferromagnetic SET as a tool to investigate the electronic structure in ferromagnetic metals 
have also been discussed.

\section*{Acknowledgment}
     This work was supported by CREST Project of Japan Science and Technology Corporation. The authors thank S. Kobayashi for fruitful discussions. They also acknowledge helpful comments on the LSDA calculation on ferromagnets by M. Yasui and H. Yamada.

\appendix

\section{Calculation of the Exchange-Correlation-Correction Factor}
In LSDA the total energy of the system without external magnetic field is written as \cite{Yamada}
\begin{equation}
E_{0}(n,m)=-Ze^{2}\sum_{\mbox{\boldmath $R$}}\int {\rm d}^{3}r\frac{n(\mbox{\boldmath $r$})}{|\mbox{\boldmath $r$}-\mbox{\boldmath $R$}|}
+\frac{e^{2}}{2}\int {\rm d}^{3}r\int {\rm d}^{3}r'\frac{n(\mbox{\boldmath $r$})n(\mbox{\boldmath $r'$})}{|\mbox{\boldmath $r$}-\mbox{\boldmath $r'$}|}
+T_{\rm s}(n,m)+E_{\rm xc}(n,m),
\end{equation}
where $n(\mbox{\boldmath $r$})$ is the electron density. The first term denotes the Coulomb energy due to the nucleus with charge $Ze$ at the site $\mbox{\boldmath $R$}$ and the second term denotes the Hartree energy. The third and fourth terms denote the kinetic energy and the exchange-correlation energy, respectively, which are functionals of the electron and spin densities, $n$ and $m$. The exchange-correlation energy $E_{\rm xc}(n,m)$ is assumed here to be known.
 From eq. (A$\cdot$1) the unperturbed Hamiltonian for an electron with spin $\sigma$ is obtained: \cite{Yamada}
\begin{equation}
{\cal H}_{\sigma}^{0}=-\frac{\hbar^{2}}{2m_{\rm e}}\nabla^{2}
+v(\mbox{\boldmath $r$}) 
+v_{\rm H}(\mbox{\boldmath $r$})
+v_{\rm xc}^{\sigma}(\mbox{\boldmath $r$}),
\end{equation}
where
\begin{eqnarray}
v(\mbox{\boldmath $r$})&=&-Ze^{2}\sum_{\mbox{\boldmath $R$}}\frac{1}{|\mbox{\boldmath $r$}-\mbox{\boldmath $R$}|},\\
v_{\rm H}(\mbox{\boldmath $r$})&=&e^{2}\int {\rm d}^{3}r'\frac{n(\mbox{\boldmath $r'$})}{|\mbox{\boldmath $r$}-\mbox{\boldmath $r'$}|},\\
v_{\rm xc}^{\sigma}(\mbox{\boldmath $r$})&=&\frac{\delta E_{\rm xc}(n,m)}{\delta n(\mbox{\boldmath $r$})}+\sigma\frac{\delta E_{\rm xc}(n,m)}{\delta m(\mbox{\boldmath $r$})}.
\end{eqnarray}
Here $\hbar$ denotes the Planck constant and $m_{\rm e}$ means the electronic mass. In order to make it more complete, a term representing the spin-orbit coupling operator should be added to the above expression of ${\cal H}_{\sigma}^{0}$. In an external magnetic field $H$, represented by the vector potential {\boldmath $A$}, the Hamiltonian may be written in the form:
\begin{equation}
{\cal H}_{\sigma}={\cal H}_{\sigma}^{0}+\frac{e^{2}}{2m_{\rm e}c^{2}}A^{2}+{\cal H}_{\sigma}'.
\end{equation}
Here ${\cal H}_{\sigma}'$ represents the enhanced perturbation and is given by
\begin{equation}
{\cal H}_{\sigma}'=-\sigma g\mu_{\rm B}H(\mbox{\boldmath{$r$}})/2+{\it \Delta}v_{\rm H}(\mbox{\boldmath{$r$}})+{\it \Delta}v_{\rm xc}^{\sigma}(\mbox{\boldmath{$r$}})-\frac{{\rm i}e\hbar}{m_{\rm e}c}\mbox{\boldmath $A$}\cdot \nabla.
\end{equation}
In this expression, the first term represents the bare spin-Zeeman effect. The second and third terms arise since magnetic field changes the local electron and spin densities $n(\mbox{\boldmath{$r$}})$ and $m(\mbox{\boldmath{$r$}})$.The last term represents the orbital term, the effect of which is discussed in Appendix B and neglected here. The induced changes in the Hartree and the exchange-correlation potentials make energies of electrons shift from those including the bare spin-Zeeman effect, and cause a correction for the chemical potential shift. 

The Helmholtz free energy of the conduction electron system is given by
\begin{equation}
F=N\zeta +{\rm Tr}\,{\it \Phi}({\cal H}),
\end{equation}
where $N$ is the number of conduction electrons in the metal, $\zeta$ is the chemical potential and
\begin{equation}
{\it \Phi}({\cal H})=-k_{\rm B}T\,{\rm ln}\,[\,1+{\rm exp}(\zeta-{\cal H})/k_{\rm B}T\,].
\end{equation}
The chemical potential is determined by the next condition so as to minimize the free energy:
\begin{equation}
\frac{\partial F}{\partial \zeta}={\rm 0}.
\end{equation}
According to the Peierls' technique, ${\it \Phi}({\cal H})$ may be expanded as follows, \cite{Peierls,KuboObata}
\begin{equation}
{\rm Tr}\,{\it \Phi}({\cal H})={\rm Tr}\,{\it \Phi}({\cal H}_{0})+\sum_{\lambda}{\it \Phi}'({\epsilon}_{\lambda})
<\lambda|{\cal H}_{\sigma}'|\lambda>+\cdots,
\end{equation}
where the representation in which ${\cal H}_{\sigma}^{0}+e^{2}A^{2}/(2m_{\rm e}c^{2})$ is diagonal is used. Based on this expansion we calculated the shift of the chemical potential between with and without ${\cal H}_{\sigma}'$. Discarding negligibly small terms, the result up to the 1st order in $H$ is as follows:
\begin{eqnarray}
{\it \Delta}\zeta_{\,\rm s}
&=&\left(\sum_{\lambda}\frac{{\rm d}f(\epsilon_{\lambda}-\zeta_{0})}{{\rm d}\epsilon_{\lambda}}\right)^{-1}\sum_{\lambda}\frac{{\rm d}f(\epsilon_{\lambda}-\zeta_{0})}{{\rm d}\epsilon_{\lambda}}<\lambda|{\cal H}_{\sigma}'|\lambda> \nonumber \\
&=&\left(\sum_{i\sigma}\frac{{\rm d}f(\epsilon_{i\sigma}-\zeta_{0})}{{\rm d}\epsilon_{i\sigma}}\right)^{-1}\sum_{i\sigma}\frac{{\rm d}f(\epsilon_{i\sigma}-\zeta_{0})}{{\rm d}\epsilon_{i\sigma}}<i\sigma|{\cal H}_{\sigma}'|i\sigma> \nonumber \\
&=&-g\mu_{\rm B}H/2\left(\sum_{i\sigma}\frac{{\rm d}f(\epsilon_{i\sigma}-\zeta_{0})}{{\rm d}\epsilon_{i\sigma}}\right)^{-1}\sum_{i\sigma}\sigma\frac{{\rm d}f(\epsilon_{i\sigma}-\zeta_{0})}{{\rm d}\epsilon_{i\sigma}} \\
&&+\left(\sum_{i\sigma}\frac{{\rm d}f(\epsilon_{i\sigma}-\zeta_{0})}{{\rm d}\epsilon_{i\sigma}}\right)^{-1}\sum_{i\sigma}\frac{{\rm d}f(\epsilon_{i\sigma}-\zeta_{0})}{{\rm d}\epsilon_{i\sigma}}<i\sigma|{\it \Delta}v_{\rm H}+{\it \Delta}v_{\rm xc}^{\sigma}|i\sigma>.
\end{eqnarray}
The second equality holds because, up to the 1st order in $H$, the eigen states of ${\cal H}_{\sigma}^{0}$ can be used in the sum. At low temperatures the expression becomes as
\begin{equation}
{\it \Delta}\zeta_{\,\rm s}=-{\cal P}g\mu_{\rm B}H/2
+\frac{1}{\rho_{+}+\rho_{-}}\sum_{i\sigma}\frac{{\rm d}f(\epsilon_{i\sigma}-\zeta_{0})}{{\rm d}\epsilon_{i\sigma}}<i\sigma|{\it \Delta}v_{\rm H}+{\it \Delta}v_{\rm xc}^{\sigma}|i\sigma>.
\end{equation}
The first term denotes the contribution by the bare spin-Zeeman effect. The second term indicates that average energy shift due to ${\it \Delta}v_{\rm H}(\mbox{\boldmath{$r$}})$ and ${\it \Delta}v_{\rm xc}^{\sigma}(\mbox{\boldmath{$r$}})$ for the electron states around the original Fermi energy composes the correction term. 
The same expression could be obtained at high magnetic field and at low temperature based on the semiclassical theory of the de-Haas van Alphen effect. \cite{LifKos}

The changes in $n(\mbox{\boldmath{$r$}})$ and $m(\mbox{\boldmath{$r$}})$, which cause ${\it \Delta}v_{\rm H}(\mbox{\boldmath{$r$}})$ and ${\it \Delta}v_{\rm xc}^{\sigma}(\mbox{\boldmath{$r$}})$, are induced by the shifts of the electron energies, and the electron energies are determined by the Hamiltonian which includes ${\it \Delta}v_{\rm H}(\mbox{\boldmath{$r$}})$ and ${\it \Delta}v_{\rm xc}^{\sigma}(\mbox{\boldmath{$r$}})$. Therefore, ${\it \Delta}v_{\rm H}(\mbox{\boldmath{$r$}})$ and ${\it \Delta}v_{\rm xc}^{\sigma}(\mbox{\boldmath{$r$}})$ should be calculated in a self-consistent way. In the homogeneous enhancement model, \cite{Yamada} induced changes in $m(\mbox{\boldmath $r$})$ and $n(\mbox{\boldmath $r$})$ by an external magnetic field with a wave-number vector, {\boldmath $q$}, are assumed to be such as
${\it \Delta}m(\mbox{\boldmath $r$})=\alpha_{\mbox{\boldmath $q$}}{\it \Delta}m_{s}(\mbox{\boldmath $r$})$ 
and
${\it \Delta}n(\mbox{\boldmath $r$})=\beta_{\mbox{\boldmath $q$}}{\it \Delta}n_{s}(\mbox{\boldmath $r$})$,
where $\alpha_{\mbox{\boldmath $q$}}$ and $\beta_{\mbox{\boldmath $q$}}$ are {\boldmath $q$}-dependent constants, and ${\it \Delta}m_{s}$ and ${\it \Delta}n_{s}$ are the unenhanced spin-density and charge-density shifts, respectively. Namely, the induced changes in $n$ and $m$ are assumed homogeneously enhanced independent of the position {\boldmath $r$}. The authors of ref. 19 solved the 1st order perturbation equations for ${\it \Delta}m(\mbox{\boldmath $r$})$ and ${\it \Delta}n(\mbox{\boldmath $r$})$ self-consistently at finite {\boldmath $q$} in LSDA. Then they calculated $\alpha_{0}$, the enhancement factor for the change in $m(\mbox{\boldmath $r$})$ in a homogeneous magnetic field, by taking the limit of {\boldmath $q$}$\rightarrow$ 0 to obtain the high-field spin susceptibility  $\chi_{\rm hf}$. By the similar calculation, $\beta_{0}$, the enhancement factor for the change in $n(\mbox{\boldmath $r$})$ in a homogeneous magnetic field, can be easily obtained:
\begin{equation}
\beta_{0}=0.
\end{equation}
This means that, in a homogeneous magnetic field, local electron density, $n(\mbox{\boldmath $r$})$, does not change. Because the Hartree potential is a functional of $n(\mbox{\boldmath $r$})$ alone, ${\it \Delta}v_{\rm H}(\mbox{\boldmath $r$})$ vanishes and one should consider only ${\it \Delta}v_{\rm xc}^{\sigma}(\mbox{\boldmath{$r$}})$ in ${\cal H}_{\sigma}'$. Moreover, because ${\it \Delta}n(\mbox{\boldmath $r$})=0$, ${\it \Delta}v_{\rm xc}^{\sigma}(\mbox{\boldmath{$r$}})$ is simplified in this case as
\begin{equation}
{\it \Delta}v_{\rm xc}^{\sigma}(\mbox{\boldmath{$r$}})=\int {\rm d}^{3}r(L(\mbox{\boldmath $r$})+\sigma J(\mbox{\boldmath $r$})){\it \Delta}m(\mbox{\boldmath $r$}),
\end{equation}
where $L(\mbox{\boldmath $r$})$ and $J(\mbox{\boldmath $r$})$ are the second functional derivatives of the exchange-correlation energy $E_{\rm xc}(n,m)$ as defined in ref. 19.
Substituting (A$\cdot$16) in (A$\cdot$14) and using $\alpha_{0}$ and $\chi_{\rm hf}$ obtained in ref. 19, expressions for the enhanced (suppressed) shift of the chemical potential (6.1) and (6.2) are obtained.

\section{Calculation of the Orbital Contribution}
Here ${\it \Delta}\zeta_{\rm orb}$, the contribution of electron orbit to the shift of the chemical potential in magnetic fields, due to the bare perturbation of the external magnetic field is calculated. The Hamiltonian representing the perturbation related to the orbit may be written in the form:
\begin{eqnarray}
{\cal H}_{\rm orb}'
&=&-\frac{{\rm i}e\hbar}{m_{\rm e}c}\mbox{\boldmath $A$}\cdot \nabla \\
&=&\mu_{\rm B}HL_{z}.
\end{eqnarray}
Here the magnetic field is assumed in $z$ direction and the usual gauge {\boldmath $A$}$=(${\boldmath $H$}$\times${\boldmath $r$}$)/2$ is used. This term adds the next term to Tr${\it \Phi}({\cal H})$ in (A$\cdot$11):
\begin{equation}
\sum_{\lambda}{\it \Phi}'(\epsilon_{\lambda})<\lambda|{\cal H}_{\rm orb}'|\lambda>.
\end{equation}
As a consequence, it causes a shift of the chemical potential of the next expression, which is additive with ${\it \Delta}\zeta_{\,\rm s}$:
\begin{equation}
{\it \Delta}\zeta_{\rm orb}=
\frac{\mu_{\rm B}H}{\rho_{+}+\rho_{-}}\sum_{\lambda}\frac{{\rm d} f({\epsilon}_{\lambda}-\zeta_{0})}{{\rm d} {\epsilon}_{\lambda}}<\lambda|{L}_{z}|\lambda>.
\end{equation}
Considering up to the 1st order in $H$, the eigen states of ${\cal H}_{\sigma}^{0}$ can be used in the above expression. Thus, we obtain eq. (6.3).


\begin{thebibliography}{99}
\bibitem{CP}The definition of the chemical potential here have arbitrariness in a sense. It should be noted, however, that the relevant thermodynamic quantity which should be defined uniquely is the {\it electrochemical} potential, and that arbitrariness of the definition of the chemical potential causes no consequence.
\bibitem{OSC}In real situations, background charges also play the role of external sources. Here, however, we neglect their effect for the sake of simplicity.
\bibitem{omegaA}As is apparent from the definition in eq. (4.4), $\omega_{\rm A}(H)$ and $\omega_{\rm A}(0)$ in eq. (4.8) represent the electrochemical potentials in the condition that A is electrostatically neutral at magnetic fields $H$ and 0. When charging effect is not negligible, tunneling-in or out of electrons makes the electrochemical potential of A shift. This effect is included in the electrostatic energy $U$ in the present formulation.
\bibitem{vector}Here, in consideration of general cases for the domain structure, we put the spatial polarity not to the spin polarization but to the magnetization, express the magnetization and the magnetic field by vectors and use the original spin polarization ${\cal P}$.
\end{thebibliography}
\end{document}